\documentclass[aip,apl,preprint,superscriptaddress]{revtex4-1}
\usepackage[T1]{fontenc}
\usepackage{times}
\usepackage{graphicx,color}
\usepackage{amsfonts,amsmath,amssymb,amsbsy}
\usepackage[colorlinks=true, linkcolor=blue, citecolor=red, urlcolor=magenta]{hyperref}

\def\blue#1{\textcolor{blue}{#1}}
\def\emph#1{\textcolor{red}{#1}}
\begin{document}

\title{Magnetic Domain Wall Engineering in a Nanoscale Permalloy Junction}

\author{Junlin Wang}
\affiliation{York-Nanjing International Center of Spintronics (YNICS), Collaborative Innovation Center of Advanced Microstructures, School of Electronic Science and Engineering, Nanjing University, Nanjing 210093, China}
\affiliation{Spintronics and Nanodevice laboratory, Department of Electronics, The University of York, York YO10 5DD, United Kingdom}

\author{Xichao Zhang}
\affiliation{School of Science and Engineering, The Chinese University of Hong Kong, Shenzhen 518067, China}

\author{Xianyang Lu}
\affiliation{Department of Physics, The University of York, York YO10 5DD, United Kingdom}

\author{Jason Zhang}
\affiliation{Department of Physics, The University of York, York YO10 5DD, United Kingdom}

\author{Yu Yan}
\affiliation{Spintronics and Nanodevice laboratory, Department of Electronics, The University of York, York YO10 5DD, United Kingdom}

\author{Hua Ling}
\affiliation{Spintronics and Nanodevice laboratory, Department of Electronics, The University of York, York YO10 5DD, United Kingdom}

\author{Jing Wu}
\affiliation{Department of Physics, The University of York, York YO10 5DD, United Kingdom}

\author{Yan Zhou}
\affiliation{School of Science and Engineering, The Chinese University of Hong Kong, Shenzhen 518067, China}

\author{Yongbing Xu}
\email{yongbing.xu@york.ac.uk}
\affiliation{York-Nanjing International Center of Spintronics (YNICS), Collaborative Innovation Center of Advanced Microstructures, School of Electronic Science and Engineering, Nanjing University, Nanjing 210093, China}
\affiliation{Spintronics and Nanodevice laboratory, Department of Electronics, The University of York, York YO10 5DD, United Kingdom}

\begin{abstract}
Nanoscale magnetic junction provides a useful approach to act as the building block for magnetoresistive random access memories (MRAM), where one of the key issues is to control the magnetic domain configuration. Here, we study the domain structure and the magnetic switching in the Permalloy (Fe$_{20}$Ni$_{80}$) nanoscale magnetic junctions with different thicknesses by using micromagnetic simulations. It is found that both the $90$-degree and $45$-degree domain walls can be formed between the junctions and the wire arms depending on the thickness of the device. The magnetic switching fields show distinct thickness dependencies with a broad peak varying from $7$ nm to $22$ nm depending on the junction sizes, and the large magnetic switching fields favor the stability of the MRAM operation.
\end{abstract}

\date{\today}
\maketitle


The magnetoresistive random access memory (MRAM)~\cite{1,2,3,4,yan2016element,liu2016transient} based on the tunneling magnetic resistance (TMR) effect has the potential to replace all existing memory devices in a computer or other hard disk drives as it could provide a high read/write operation speed and is also nonvolatile~\cite{5,6,7}. On the other hand, the magnetic domain wall gives a flexible approach in the data storage as well as the logic circuit~\cite{8,9,10,11}. Comparing with the TMR-based MRAM devices, a domain wall motion-based magnetic junction could have a single layer structure, which might have great advantages in terms of fabrication and application~\cite{12,13}. The magnetic junction shows several types of the magnetoresistance effect by applying magnetic field~\cite{14,15}.

The magnetic switching in the junction structure can be controlled by either the external magnetic field or the applied electrical current~\cite{16,17}. Recently, it has been reported that the magnetic switching induced by the spin-transfer torque (STT) can enable the junction to work as a STT-MRAM device~\cite{18,19}. There are also reports about the spin-polarized current can induce the junction to generate spin waves~\cite{20,21,22}. It is found that the Permalloy junction has several metastable magnetization states, which can be used to store the information~\cite{16}. Thus, reliable control of the magnetic domain configuration in the magnetic junction is an important task. In this \textit{Letter}, we present a micromagnetic study of the domain structures and the magnetic switching in the nanoscale Permalloy junctions within magnetic cross structures with different thicknesses. The numerical simulations are carried out by using the Object Oriented MicroMagnetic Framework (OOMMF) software~\cite{23}. It is found that the junction thickness has distinct effects on the domain wall configuration, the initial magnetic switching, as well as the coercivity. In the initial states, both $45$-degree and $90$-degree domain walls are found to be formed in the studied model. Both the initial magnetic switching and the coercivity show nonlinear dependences on the thickness, indicating the importance of controlling the thickness for the writing process when the nanoscale Permalloy junction works as a building block in information storage devices.


The micromagnetic simulations are performed using the standard micromagnetic simulator OOMMF software~\cite{23}, which stands on the Landau-Lifshitz-Gilbert equation~\cite{24,25},
\begin{equation}
\frac{d\boldsymbol{M}}{dt}=-\left|\gamma\right|\boldsymbol{M}\times\boldsymbol{H}_{\text{eff}}+\frac{\alpha}{M_{\text{S}}}\left(\boldsymbol{M}\times\frac{d\boldsymbol{M}}{dt}\right),
\label{LLG}
\end{equation}
where $\boldsymbol{M}$ is the magnetization of the magnetic layer, $M_{\text{S}}$ is the saturation magnetization, $\gamma$ is the Gilbert gyromagnetic ratio and $\alpha$ is the damping constant. $\boldsymbol{H}_{\text{eff}}$ is the effective field, which is derived from the magnetic energy density,
\begin{equation}
\boldsymbol{H}_{\text{eff}}=-\mu_{0}^{-1}\frac{\delta\varepsilon}{\delta\boldsymbol{M}},
\label{energy-density}
\end{equation}
where $\varepsilon$ contains the Heisenberg exchange, anisotropy, applied magnetic field, and demagnetization energy terms.

The magnetic material used for the micromagnetic simulation is Permalloy, \textit{i.e}., Fe$_{20}$Ni$_{80}$ alloy, which has a low coercivity and a high permeability. The size of the cross structure of the nanoscale junction is defined as $10$ nm $\times$ $10$ nm with a varying thickness from $2.5$ nm to $25$ nm and the length of all the wire arms is fixed at $200$ nm. The saturation magnetization is equal to $8.6\times 10^{5}$ A/m. The exchange stiffness and the crystalline anisotropy constant are set at $13\times 10^{-12}$ J/m and $0$ J/m$^{3}$, respectively. The simulation cell size is set as $2.5$ nm $\times$ $2.5$ nm $\times$ $2.5$ nm, which is compared with the exchange length ($5.3$ nm) of Permalloy.

\begin{figure}[t]
\centerline{\includegraphics[width=0.7\textwidth]{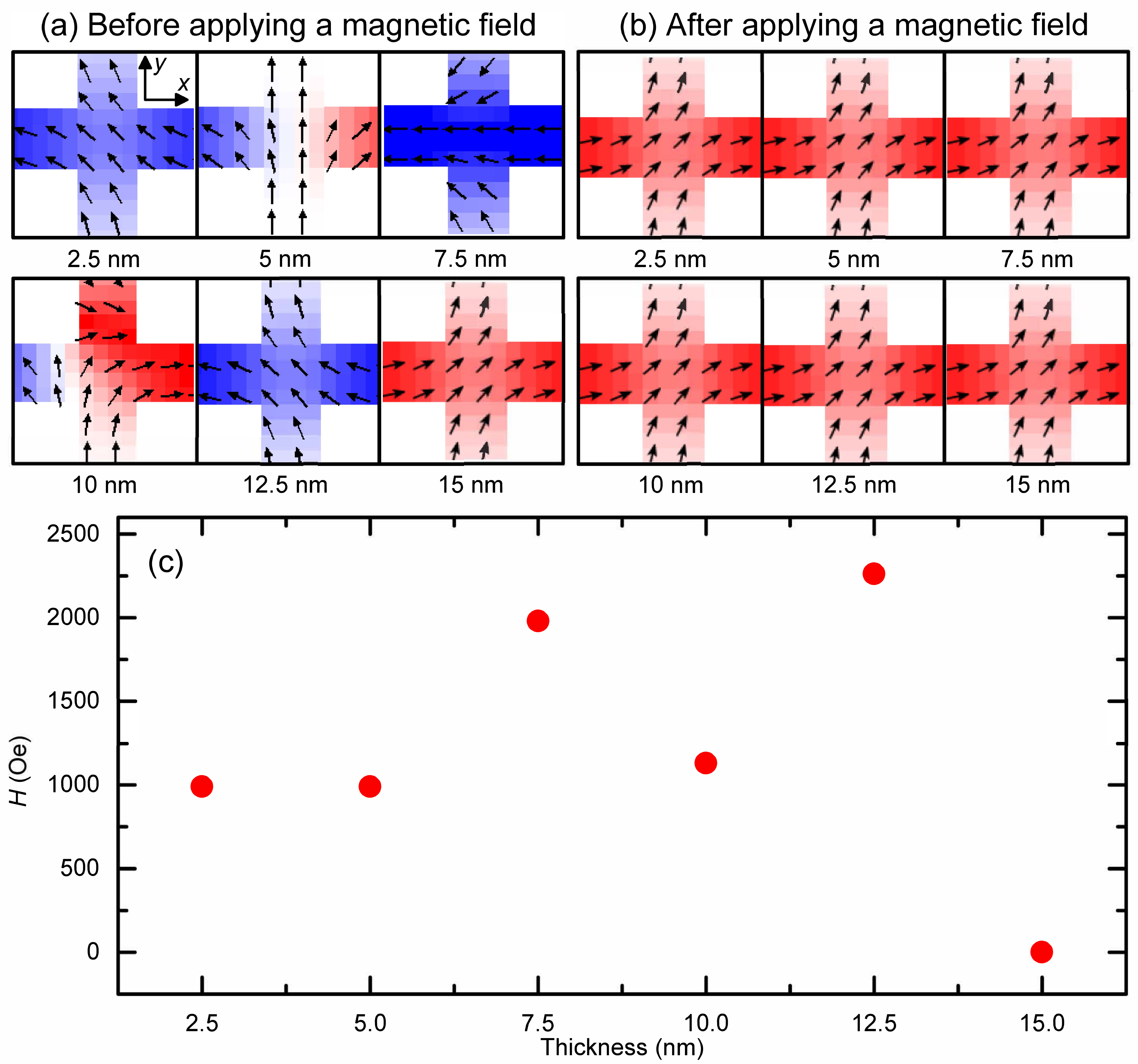}}
\caption{%
(a)The magnetic domain configurations in the cross structure relaxed from random magnetization distributions for different thicknesses. (b) The magnetic domain configurations obtained by applying a magnetic field. (c) The amplitude of the magnetic field which changes the domain configurations shown in (a) to that shown in (b).
}
\label{FIG1}
\end{figure}

The initially relaxed magnetization distributions around the junctions of the cross structures with different thicknesses are shown in Fig.~\ref{FIG1}(a), which are obtained by relaxing the cross structures with random magnetization distributions. The spins in the wire arms are all aligned in parallel along the wire directions due to the strong shape anisotropy. There are three types of domain configurations around the junctions. For the junctions with the thicknesses of $2.5$ nm, $7.5$ nm and $12.5$ nm, the spins in the junction are aligned in parallel with the spins in one of the wires, and the $90$-degree domain walls form between another wires. For the cases with the thicknesses of $5$ nm and $15$ nm, the spins in the junctions are aligned largely in parallel, which can be described as a single magnetic domain or a coherent spin block (CSB), and the spin direction of these CSBs is in $45$ degree with those in both wires. For the cases with thickness of $10$ nm, the spins in the junction are form a $90$-degree domain wall within the junction. Indeed, the initial magnetization distribution in the cross structure can also be controlled by the applying an external magnetic field. As shown in Fig.~\ref{FIG1}(b), the initial magnetization distribution in the cross structure can be modified to be $45$-degree domain by applying a magnetic field pointing at an angle of $45$ degree to the $+x$-direction. The required amplitude of the magnetic field corresponding to different thickness is given in Fig.~\ref{FIG1}(c). The applied magnetic field changes the domain structure to the coherence switching mode from the initially relaxed states, where the required magnetic field is different for samples with different initially relaxed magnetization distribution and thickness. Note that the magnetization distribution configuration in the thickness direction is uniform (\blue{see Supplementary Information Fig. S1}).

\begin{figure}[t]
\centerline{\includegraphics[width=0.7\textwidth]{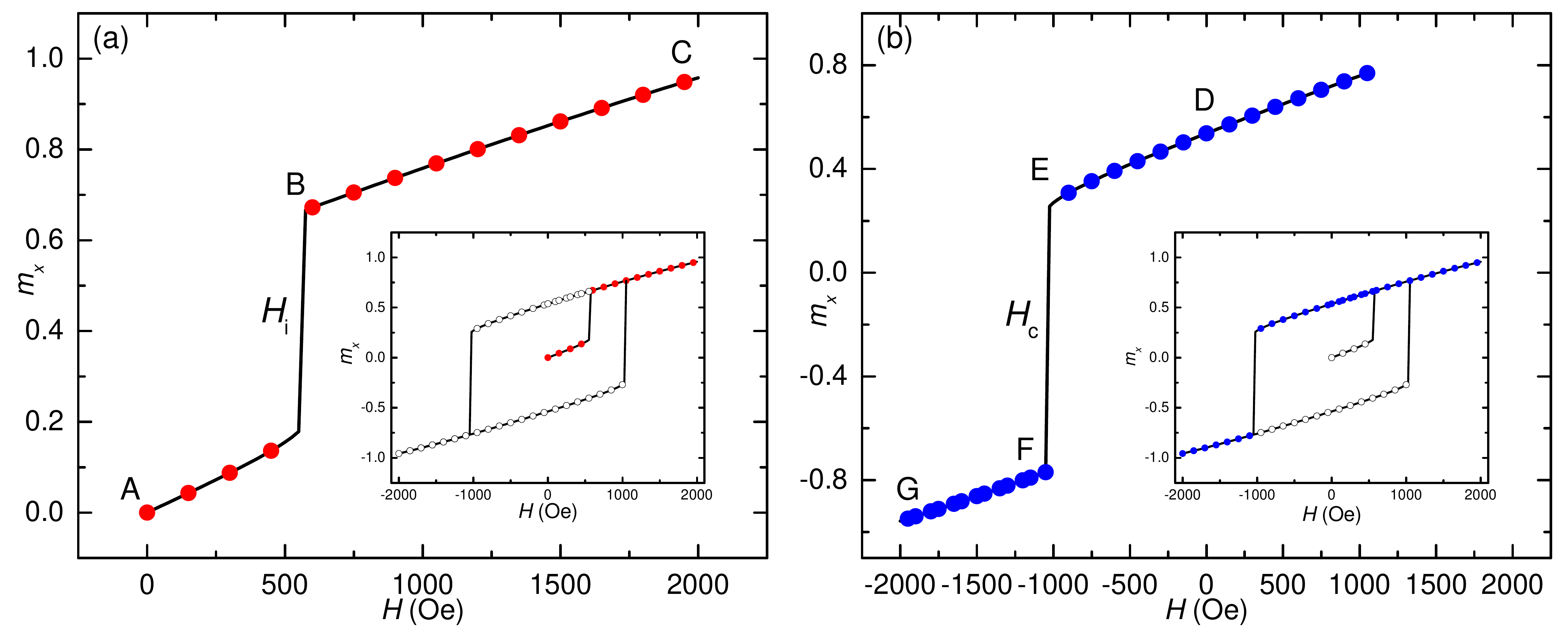}}
\caption{%
The hysteresis loop for the cross structure with the thickness of $2.5$ nm. The initial switching magnetic field ($H_{\text{i}}$) and the coherence switching magnetic field ($H_{\text{c}}$) are indicated in (a) and (b), respectively.
}
\label{FIG2}
\end{figure}
\begin{figure}[t]
\centerline{\includegraphics[width=0.7\textwidth]{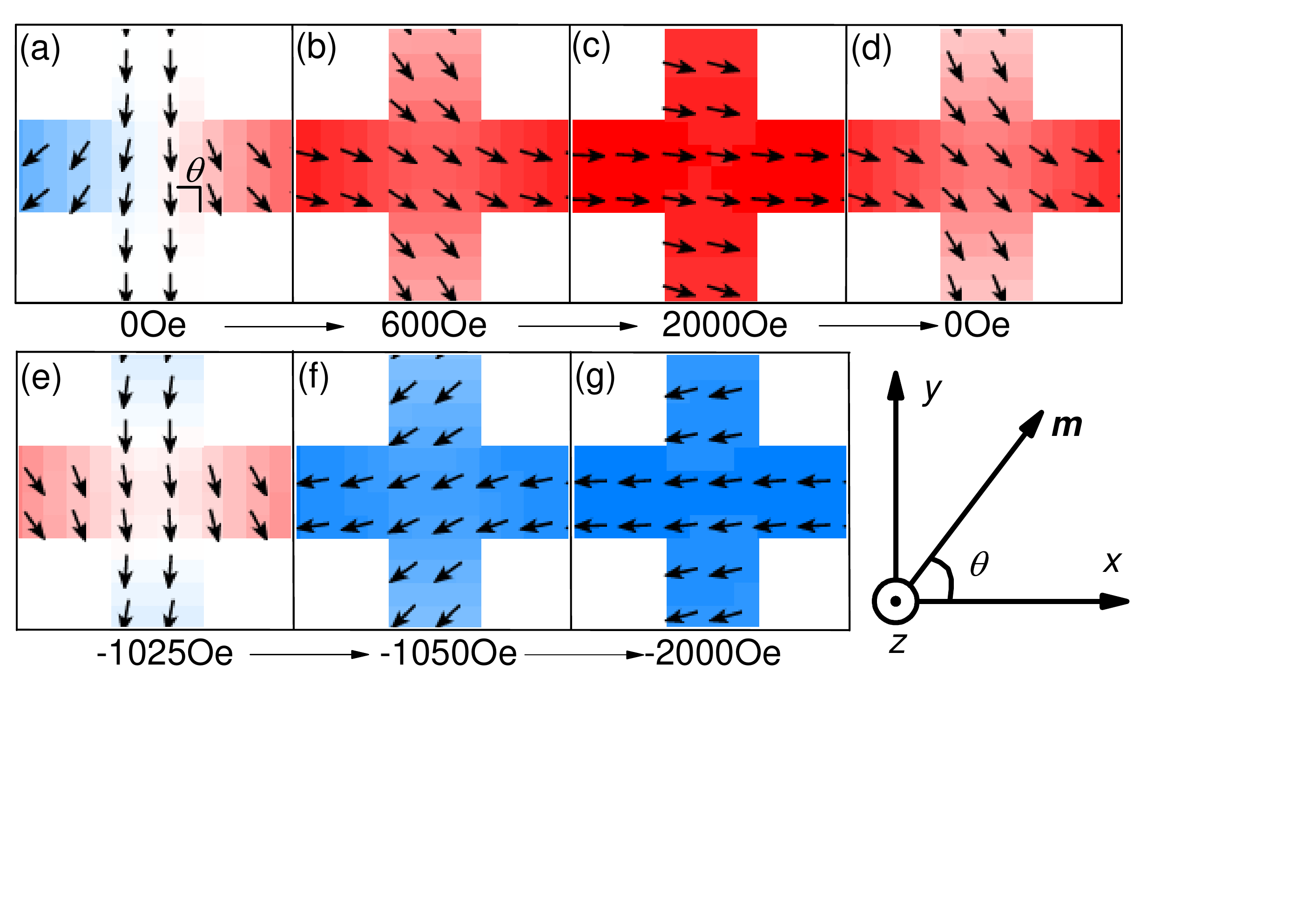}}
\caption{%
The magnetic domain configurations of the $2.5$-nm-thick junction at different applied magnetic fields. The magnetic field is applied along the $x$-direction. The labels correspond to the states indicated in Fig.~\ref{FIG2}, \textit{i.e}., (a) $H=0$ Oe, (b) $H=600$ Oe (before switching), (c) $H=2000$ Oe (after switching), (d) $H=0$ Oe, (e) $H=-1025$ Oe (before switching), (f) $H=-1050$ Oe (after switching), and (g) $H=-2000$ Oe.
}
\label{FIG3}
\end{figure}

In the following, we study the magnetization switching process driven by an external magnetic field for the junctions with different thicknesses. The simulated hysteresis loops are the same in different layers of the device. Fig.~\ref{FIG2} shows the result for the $2.5$-nm-thick junction. The magnetic field is first applied along the $+x$-direction, of which the amplitude first increases from $0$ Oe to $2000$ Oe and reduces to $0$ Oe. Then, the magnetic field changes in the same manner but along the $-x$-direction. The simulated hysteresis loop is given in Fig.~\ref{FIG2} and the magnetization configurations illustrated in Fig.~\ref{FIG3} are corresponding to the marked states in the hysteresis loop given in Fig.~\ref{FIG2}, which represent the magnetic switching process in the nanoscale junction. The initial magnetization configuration in the nanoscale junction is given in Fig.~\ref{FIG3}. The magnetic field of the first magnetic switching from the initially relaxed state to the state with a $45$-degree domain wall is defined as the initial magnetization switching field ($H_{\text{i}}$), which is indicated in Fig.~\ref{FIG2}(a). As the applied magnetic field increases from $0$ Oe to $2000$ Oe, the direction of the magnetization in the junction is switched where the amplitude of the critical switching field, \textit{i.e}., the coercivity ($H_{\text{c}}$) of the junction as indicated in Fig.~\ref{FIG2}(b), is equal to $1050$ Oe.

The snapshots of the switching process given in Fig.~\ref{FIG3} further show that the switching of the magnetization in the junction is coherent. The angle between the $x$-axis and the spins at the junction is defined as $\theta$. Before applying the magnetic field, the spins in the junction are in parallel with the $y$-axis and the $\theta$ is equal to $-90$ degree. By increasing the applied magnetic field above $2000$ Oe, $\theta$ increases and then reaches $0$ degree. When the applied magnetic field is reduced to $0$ Oe, $\theta$ decreases to be $-45$ degree. That means in the remanence state, the spins in the junction are aligned $45$ degree away from the wire direction, and $45$-degree domain walls are formed between the junction and the wires. When the applied magnetic field decreases from $0$ Oe to $-2000$ Oe in the $x$-direction, the spins change to be paralleled with the $x$-direction and the $\theta$ increases from $-45$ degree to $0$ degree. It is found that the whole magnetization switching process in the junction is coherent and reversible.

\begin{figure}[t]
\centerline{\includegraphics[width=0.7\textwidth]{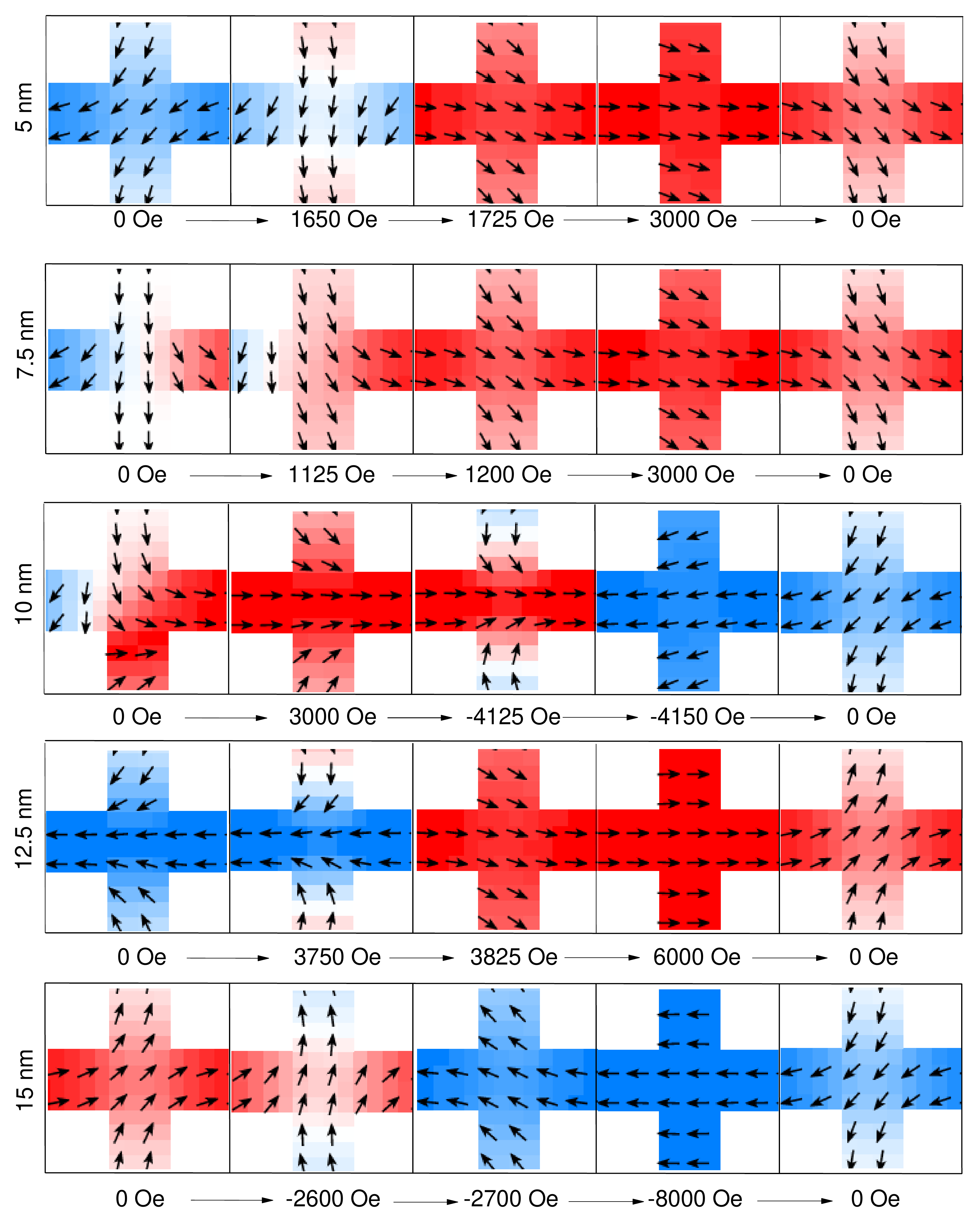}}
\caption{%
The magnetic domain configurations of the junctions at different applied magnetic fields for different thicknesses. The magnetic field is applied along the $x$-direction.
}
\label{FIG4}
\end{figure}

The typical spin configurations during the magnetic switching process of the junctions with different thicknesses are shown in Fig.~\ref{FIG4}. From the Fig.~\ref{FIG1}(a), we found that the junction with a different thickness usually has a different relaxed state. Before using the magnetic field to achieve reversible magnetic switching in the cross structure, the spins in the cross structure have to be tuned to the coherent switching mode. The coherent switching mode is defined as the states in Fig.~\ref{FIG3}(b) which shows $45$-degree domain walls. The processes to enable the coherence switching modes are shown in Fig.~\ref{FIG4}. Unlike the magnetic switching processes shown in Fig.~\ref{FIG3}, the relaxed magnetization configuration at $0$ Oe is different from the junctions with the thicknesses of $5$ nm, $7.5$ nm, $10$ nm, $12.5$ nm and $15$ nm. The spin configurations of the junctions with the thicknesses of $5$ nm and $15$ nm are similar to the spin configurations in Fig.~\ref{FIG3}(d) which are in the coherent switching mode. The junction with a thickness of $7.5$ nm has the parallel spins in the cross structure with a $\theta$ as $-90$ degree between the $x$-direction. As the magnetic field increases to $3000$ Oe, $\theta$ decreases to $0$ degree in the junction. When the magnetic field reduces back to $0$ Oe, $\theta$ increases to $45$ degree and the CSB is formed. From the junction with a thickness of $10$ nm, the initial $90$-degree DW within the junction can be eliminated and the CSB can be formed by controlling the magnetization process. When the thickness of the junction is $12.5$ nm, the spins in the cross structure requires a large magnetic field up to $3825$ Oe to reverse the magnetization direction. For the whole cross structures with the thicknesses of $5$ nm, $7.5$ nm and $10$ nm, the spin configurations in the $y$-arms have not been changed and the spin configuration in the $x$-arms can be switched. However, for the cross structures with thicknesses of $12.5$ nm and $15$ nm, the spin directions rotate along the $y$-axis as well.

\begin{figure}[t]
\centerline{\includegraphics[width=0.7\textwidth]{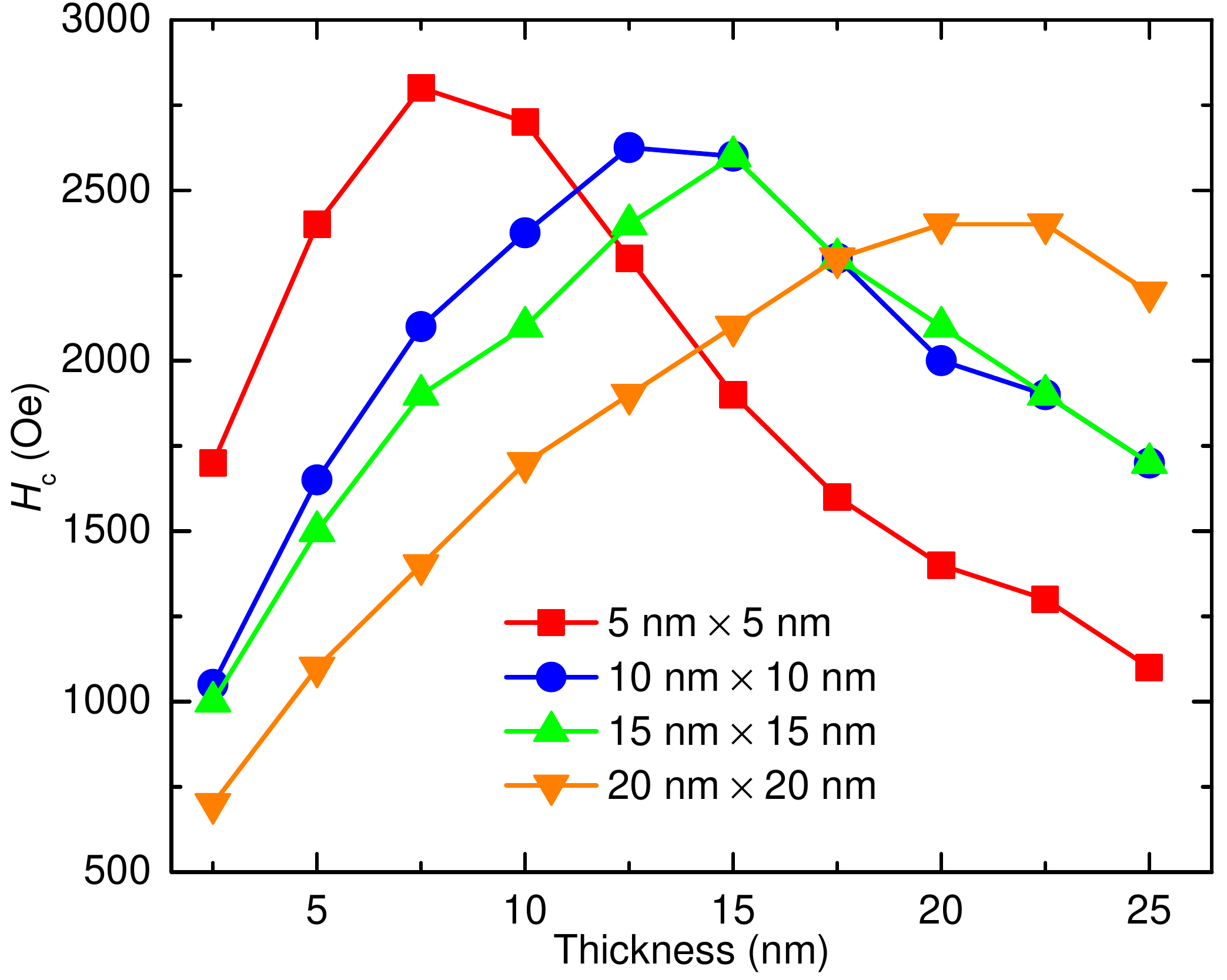}}
\caption{%
The coercivity $H_{\text{c}}$, i.e., the switching field, as a function of the thickness for the junctions with different cross section sizes.
}
\label{FIG5}
\end{figure}

The magnetic switching field of the junction can be affected by the thickness of the junction. Fig.~\ref{FIG5} shows $H_{\text{c}}$ as a function of the thickness for the junctions with different lateral sizes. The size of the device is varying from $100$ nm $\times$ $100$ nm to $400$ nm $\times$ $400$ nm, \textit{i.e}., the length of the cross structure is varying from $5$ nm to $20$ nm. $H_{\text{c}}$ increases first with increasing thickness and then decreases, showing a broad peak from $7$ nm to $22$ nm depending on the junction sizes. The reason is that, for the junction with a certain cross section size, the magnetization switching is coherent. Thus, $H_{\text{c}}$ is proportional to the total magnetization, which also means $H_{\text{c}}$ increases with the thickness as the total magnetization is proportional to the thickness. However, when the thickness is larger than a certain critical value, multiple domains can be formed during the magnetization switching, leading to incoherent magnetization switching. In such a case, $H_{\text{c}}$ decreases with the thickness, as thicker junction is more likely to form multiple domains, due to the demagnetization effect. Besides, the critical value of the thickness increases with increasing lateral dimensions of the junction. The reason is that the amplitude of the demagnetization effect resulting in the incoherent switching is proportional to thickness and is inversely proportional to the lateral dimensions of the junction (\blue{see Supplementary Information Fig. S2-S4}). We note that there are large magnetic switching fields of around $2500$ Oe, which favor the stability of the MRAM operation. The large magnetic switching fields, however, may lead to large current needed for the switching of the junction. One may need to explore the spin torque transfer or thermal assistant switching for future applications.

In conclusion, we have carried out a full micromagnetic study on the magnetization configuration as well as the magnetic switching process in a nanoscale Permalloy junction. The relationship between the magnetic switching fields and the thickness of the nanoscale junction has been investigated. While different types of domain walls can be formed in the initially relaxed states depending on the specific thicknesses, the junction acts as a single CSB where the spins are aligned in parallel during the magnetization process. The magnetization direction can be controlled and switched coherently by applying an external magnetic field. Both the initial magnetization field and the coercivity are found to depend on the thickness, and the large coercivity could enhance the stability of the device operation. Our work shows that the nanoscale magnetic junction has the potential to be used as a building block for future spin-based data storage or logic computing technologies.

\section*{Supplementary Material}
See supplementary material for magnetization distribution in device at different layers and side views of the magnetization distribution in device.
\section*{Acknowledgment}

This work was supported by State Key Program for Basic Research of China (Grant No.  2014CB921101, 2016YFA0300803 ), NSFC (Grants No. 61427812, 11574137), Jiangsu NSF (BK20140054), Jiangsu Shuangchuang Team Program, Shenzhen Fundamental Research Fund under Grant No. JCYJ20160331164412545 and the UK EPSRC (EP/G010064/1). 

\bibliography{References}

\newpage

\clearpage

\clearpage

\clearpage

\clearpage

\end{document}